\documentclass[usenatbib]{mnras}
\usepackage[T1]{fontenc}
\usepackage{ae,aecompl}
\usepackage{bm}
\usepackage{amsmath,amssymb}
\usepackage{dcolumn}
\usepackage[dvips]{graphicx}
\usepackage{epsf}
\usepackage{color}
\usepackage{epsf}
\usepackage{epsfig}
\usepackage{graphicx}
\usepackage{longtable}
\usepackage{psfrag}
\usepackage{txfonts}

\title{Is there evidence for anomalous dipole anisotropy in the large-scale structure?}

\author[C. A. P. Bengaly Jr. et al.]{
C. A. P. Bengaly Jr.,$^{1}$\thanks{E-mail: carlosap@on.br}
A. Bernui,$^{1}$\thanks{E-mail: bernui@on.br}
J. S. Alcaniz,$^{1}$\thanks{E-mail: alcaniz@on.br}
H. S. Xavier$^{2}$\thanks{E-mail: hsxavier@if.usp.br}
C. P. Novaes$^{1}$\thanks{E-mail: camilanovaes@on.br}
\\
$^{1}$Observat\'orio Nacional, Rua General Jos\'e Cristino, 77, S\~ao Crist\'ov\~ao, 20921-400, Rio de Janeiro - RJ, Brasil,\\
$^{2}$Instituto de Astronomia, Geof\'isica e Ci\^encias Atmosf\'ericas, Universidade de S\~ao Paulo, Rua do Mat\~ao, 1226, 05508-090, S\~ao Paulo - SP, Brasil\\
}

\date{\today}

\begin{document}
\label{firstpage}
\pagerange{\pageref{firstpage}--\pageref{lastpage}}
\maketitle

\begin{abstract}
We probe the anisotropy of the large-scale structure (LSS) with the WISE-2MASS catalogue. 
This analysis is performed by a directional comparison of the galaxy number counts through the entire celestial sphere once systematic effects, such as star-galaxy separation and foregrounds contamination, are properly taken into account. 
We find a maximal hemispherical asymmetry whose dipolar component is $A = 0.0507 \pm 0.0014$ toward the $(l,b) = (323^{\circ},-5^{\circ})$ direction, whose result is consistent with previous estimations of our proper motion in low and intermediate redshifts, as those carried out with Type Ia Supernovae and similar LSS catalogues.
Furthermore, this dipole amplitude is statistically consistent ($p$-value = $0.061$) with mock catalogues simulated according to the expected $\Lambda$CDM matter density fluctuations, in addition to observational biases such as the incomplete celestial coverage and anisotropic sky exposure. 
Our results suggest, therefore, that there is no strong evidence for anomalous anisotropy in the LSS, given the limitations and systematics of current data, in the concordance model scenario. 
\end{abstract}

\begin{keywords}
Cosmology: observations; The large-scale structure of the Universe; Cosmological Principle
\end{keywords}

\section{Introduction} \label{intro}

The isotropy of the Universe is one of the most fundamental pillars of the standard model of cosmology. Along with the hypothesis that the Universe must be homogeneous on large scales, it constitutes the so-called Cosmological Principle (CP), which states that there are no intrinsic privileged directions and positions throughout its entire extension~\citep{goodman95, wu99, maartens11, clarkson12}. Albeit the standard $\Lambda$CDM model, which is based on the CP, excellently describes cosmological observations such as the Cosmic Microwave Background (CMB) temperature fluctuations~\citep{hinshaw13, planck15a}, the growth rate of the large-scale structure (LSS)~\citep{peacock01, blake11a, reid12}, besides time scales from the galaxy ages, and cosmological distances from Type Ia Supernovae (SNe) as well as baryonic acoustic oscillations (BAO)~\citep{riess98, p99, alcaniz99, alcaniz03, eisenstein05, blake11b, beutler11, suzuki12, betoule14, boss15, gabriela16}, the validity of the CP remains yet to be directly assessed. Therefore, it is crucial to perform observational tests of cosmic isotropy and homogeneity, since a violation of at least one of such hypotheses would lead to a complete reformulation of the standard cosmological scenario. 

It has been well known that the CMB exhibits a dipolar anisotropy in its temperature due to Doppler boost and aberration effects, i.e., the kinematic dipole, which is attributed to our relative motion through the Universe. Such effect was predicted and detected still in the late 1960s~\citep{stewart67, conklin69}, being confirmed many years later in the full-sky CMB map from COBE~\citep{kogut93}, WMAP~\citep{hinshaw09} and Planck~\citep{planck14} maps. This measurement shows that our motion is characterised by a velocity of $\mbox{\tt v} \simeq 384 \; km/s$ toward the $(l,b) = (264^{\circ},48^{\circ})$ direction. Attempts to obtain this signal in the LSS appeared as early as the 1960s~\citep{devaucouleurs68}, whose efforts increased given the rapidly growing improvements of infra-red and optical galaxy surveys, such as the IRAS PSCz and 2MASS, during the following decades~\citep{yahil86, lahav88, lynden-bell89, strauss92, basilakos98, rowan-robinson00, maller03, erdogdu06, basilakos06, bilicki11}. The goal, in these cases, was the amplitude and direction of the so-called clustering dipole, which gives a preferred direction in the low-$z$ LSS due to the galaxy clustering responsible for the acceleration of the Local Group, thus allowing one to determine the consistency (and convergence) with the CMB kinematic dipole. 

One of the most popular approaches to perform the estimation of the clustering dipole makes use of galaxy luminosity function and magnitudes measured in different distances (thus the flux-weighted dipole). Then, this information is compared with the expected peculiar velocity in these corresponding scales via linear perturbation theory~\citep{peebles80}. Some of the latest results reported dipoles roughly aligned with the CMB one. For example,~\cite{rowan-robinson00} obtained $(l,b) = (267^{\circ}, 50^{\circ})$ adopting IRAS PSCz data set, while~\cite{maller03} and~\cite{erdogdu06} analyses provided $(l,b) = (264.5^{\circ}, 43.5^{\circ})$ and $(l,b) = (245^{\circ}, 39^{\circ})$, respectively, both using different versions of the 2MASS data. The convergence of this flux-weighted dipole amplitude in large scales, nevertheless, is still a matter of debate~\citep{bilicki11}. Similar analyses have been carried out in the X-ray spectrum with this flux-weighted method in the galaxy clusters luminosity function~\citep{plionis98, kocevski04}, besides the dipole anisotropy in the diffuse X-ray background due to Compton-Getting effect with both ROSAT~\citep{plionis99} and HEAO1~\citep{scharf00, boughn02} surveys. The results are similar to the optical and infra-red analyses, although many authors claimed greater difficulties in the X-ray band than in the former cases because of less controlled systematics. 

Given the increasing amount and precision of the observational LSS data, it has become possible to probe its dipole anisotropy using only projected two-dimensional counts from these all-sky galaxy catalogues\footnote{Some attempts on performing such analyses in partial sky maps can be found in~\cite{pullen10, hazra15} for estimates on Sloan Digital Sky Survey (SDSS) galaxy and Lyman-$\alpha$, respectively, besides the LSS dipole test applied in SDSS galaxy sample from Data Release~6~\citep{itoh10}.}: \cite{gibelyou12} performed such analysis with a large variety of observational samples,~\cite{appleby14} and~\cite{alonso15} adopted the 2MASS data with photometric redshifts~\citep[2MPZ,][]{bilicki14} with this purpose, whereas~\cite{yoon14} performed a dipole estimation with the Wide Infrared Satellite Explorer~\citep[WISE,][]{wright10} data. All these works presented good agreement with the clustering dipole results obtained with the flux-weighted method\footnote{We remark that we had only considered the two-dimensional projected approach in our analyses, not the flux-weighted one. A detailed discussion on these approaches, in addition to the expected consistency between each other, and the behaviour of the dipole amplitude when reaching deeper and deeper scales is made in~\cite{gibelyou12}}, with the advantage that this 2D projected estimators do not require further information (or assumptions) about the magnitudes and luminosity function of the sources other than their celestial distribution.  

On the other hand, the isotropy of the large-scale LSS had been tested with radio surveys as well, which are much deeper than the current optical and infra-red catalogues, thus, being ideal data sets to directly probe the consistency with the CMB kinematic dipole. One of the early attempts were carried out by~\cite{baleisis98}, who joined both Green Bank and Parkes-MIT-NRAO radio catalogues, yet could not detect such signal due to the large shot-noise number counts in this sample. The first statistical significant report of this dipole were made by~\cite{blake02}, who adopted the much larger radio catalogue from NVSS and obtained a velocity dipole whose amplitude and direction shows reasonably good agreement with the CMB's. However~\cite{singal11} revised these data and found a nearly five times larger velocity than the expected value from the CMB dipole, although the direction is still roughly consistent with it, as well as~\cite{blake02} results. This surprising result was later confirmed by~\cite{rubart13, tiwari14, cobos14, tiwari15}, albeit with moderately smaller velocities, while later on~\cite{tiwari16} showed that a bias choice of $b \geq 2.0$ for these radio sources decreases this discrepancy between the CMB and the radio dipole to $2.3\sigma - 2.8\sigma$. In addition, some previous works obtained large scale flows using peculiar velocity probes, such as nearby galaxies and SNe~\citep{watkins09} as well as the kinematic Sunyaev-Zeldovich (kSZ) effect from galaxy clusters~\citep{kashlinsky09, kashlinsky10, kashlinsky11, atrio15}, than the $\Lambda$CDM predictions in the scales they probed, besides claims of statistical isotropy violation in the CMB temperature fluctuations, e.g. the hemispherical power asymmetry firstly showed in the first WMAP release~\cite{eriksen04, hansen04}, then confirmed in its later versions, as well as in the Planck data~\citep{b08, bop14, hoftuft09, akrami14, polastri15, zhao15, ghosh16, cheng16}\footnote{For the interested reader on other CMB features, such as large-scale alignments and power deficit, we refer to~\cite{schwarz15}.}. The physical motivation of these possible violations of isotropy hypothesis remains an open issue by the present moment.

In the light of these intriguing results, it is of great importance to test the isotropy assumption with other probes and methods, investigating whether a similar feature can be detected in these data. Therefore, we adopt the WISE-2MASS catalogue in this work. Our goal is to look for large-angle anisotropies in the galaxy number counts (GNC) caused by galaxy clustering, and to determine the statistical significance of the result according to the expectations of the concordance model. This assessment is performed via log-normal mock realisations, which incorporate the matter density fluctuations expected from the $\Lambda$CDM model given the WISE-2MASS source counts variance, besides potential observational bias such as incomplete sky coverage, because of dust obscurity, as well as anisotropic sky exposure of WISE's observational strategy. Also, it may provide a consistency check with previous CMB and LSS anisotropy studies, besides complimentary analyses performed with cosmological distance indicators such as SNe or nearby galaxies~\citep{watkins09, antoniou10, colin11, dai11, mariano12, turnbull12, wiltshire13, cai13, rathaus13, kalus13, feindt13, ma13, jimenez15, appleby15, bengaly15a, javanmardi15, carvalho15, springbob16, lin16a, mckay16, migkas16, carvalho16, lin16b, bolejko16}. 

Hence, the structure of the paper is the following: In Section 2 we describe the WISE data, the construction of its catalogue, and the appropriate masking procedure in order to avoid foreground contamination. Section 3 is dedicated to the methodology developed for testing the GNC anisotropy, as well as the construction of the mock data. The obtained results are presented in Section 4, followed by a discussion and our main conclusions in Section 5.  

\vspace{1.5cm}

\section{Data set preparation} \label{ds-prep}

\begin{figure*}
\includegraphics[width=6.3cm, height=8.5cm, angle=90]{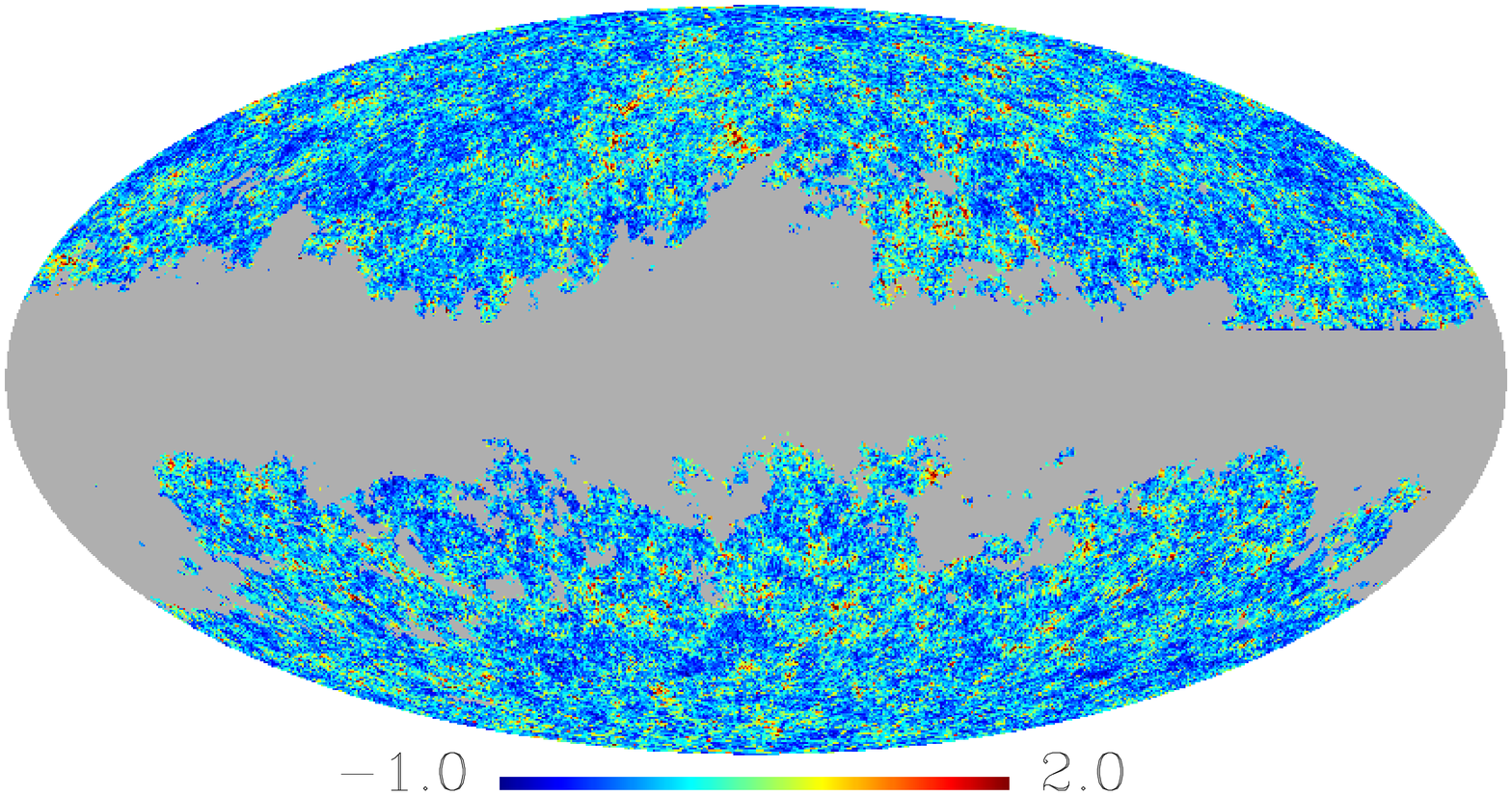}
\includegraphics[width=6.3cm, height=8.5cm, angle=90]{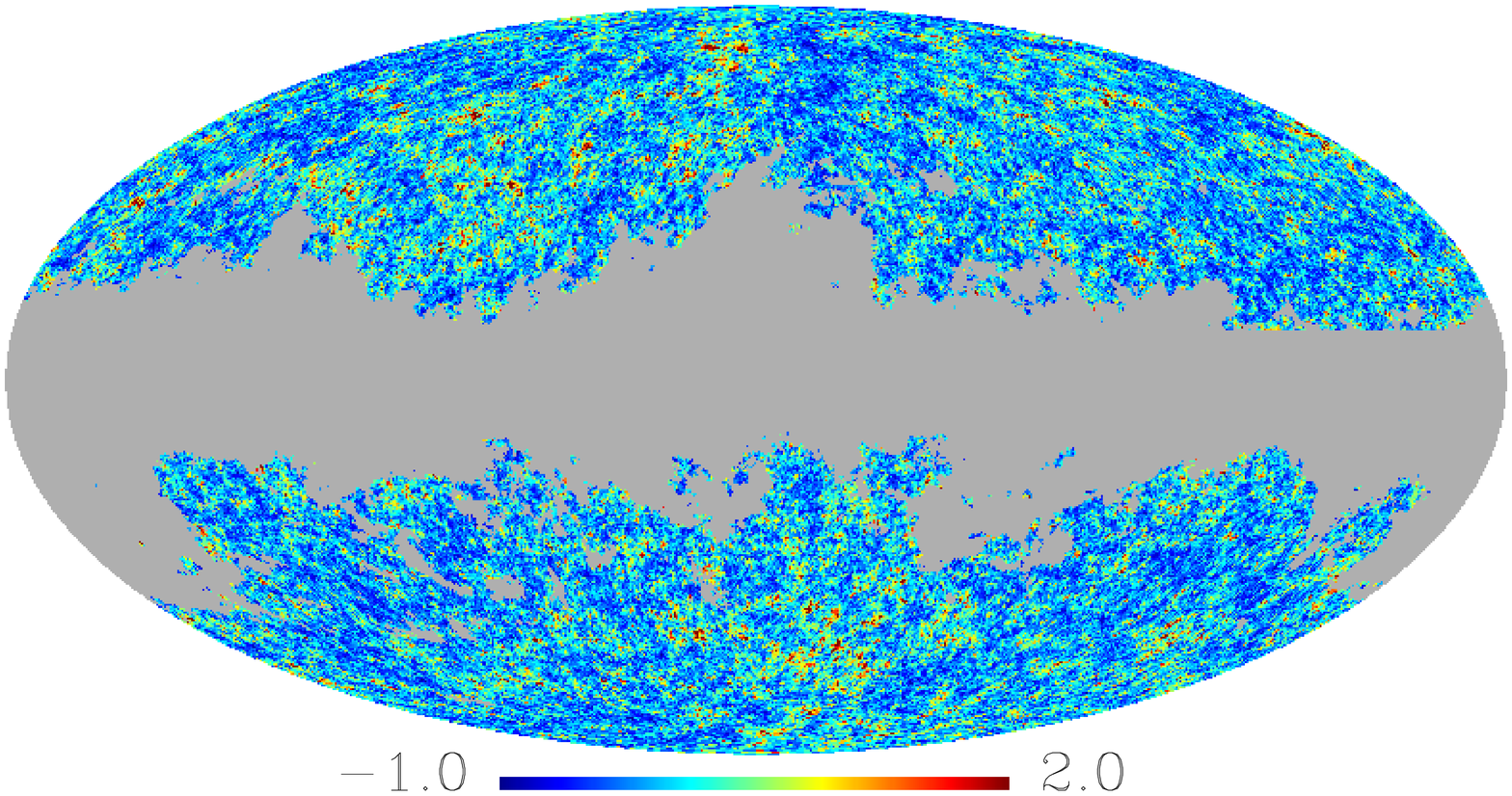}
\caption{{\it Left panel}: Mollweide projection of the WISE-2MASS sources after the colour and magnitude cuts, besides the masking template, have been properly applied. We show the number density contrast of galaxies encompassed in each pixel defined by the $N_{side}=128$ HEALpix grid, whose average number of sources per pixel is roughly $14$. {\it Right panel}: Mollweide projection of a mock WISE-2MASS map created using lognormal realisations of galaxy density fields in an isotropic $\Lambda$CDM model as observed by the WISE survey. The colour scale has been truncated at a maximal of $2.0$ in both maps in order to ease visualisation of this number density contrast.}
\label{fig:wise_sim_map}
\end{figure*}

The data set adopted in this analysis corresponds to the WISE release named "AllWISE", which is publicly available since late 2013 in the {\sc IRSA} website\footnote{\url{http://irsa.ipac.caltech.edu}}. It originally consists of nearly 750 million objects in four bands centred at $3.4, \; 4.6, \; 12, \;$ and $22 \; \mathrm{\mu m}$ wavelengths, corresponding to the $W_1$, $W_2$, $W_3$, and $W_4$. We follow the points discussed by~\cite{kovacs15a}, who matched WISE and 2MASS magnitude in the $J$ band (henceforth $J_{2MASS}$ in order to enhance the WISE biases and purity, such as removing objects according lying in the $12.0 < W_1 < 15.2$ range, besides the $J_{2MASS} > 16.5$ cut, as these authors showed that such objects are spatially biased due to the instrumental capability limitations in the latter case. In addition, a colour cut of $W_1 - J_{2MASS} < -1.7$ has been carried out in order to optimise the galaxy-star separation (hence the name WISE-2MASS of the catalogue). From the original value of 750 million objects, nearly 2.4 million points remained after applying these criteria. 

Another crucial issue in data sets like WISE-2MASS is Galactic contamination. Therefore, a mask template has been constructed following~\cite{kovacs15a} as well, where, in practice, we have removed pixels in the sky for which the colour excess satisfies $E(B-V) \geq 0.1$ according to the reddening map provided by~\cite{schlegel98}\footnote{Downloaded from the {\sc Lambda} website\url{http://lambda.gsfc.nasa.gov}}. Pixels localised in regions with high extinction contamination according to the WMAP dust template, obtained in the same web site, have also been removed, albeit most of them coincide with the reddening constraint. The resulting map of extra-galactic sources is shown in the left panel of Figure~\ref{fig:wise_sim_map} with HEALpix~\citep{gorski05} $N_{side}=128$ resolution. It comprises $\sim 1.7$ million sources with observed sky fraction of $f_{sky} \simeq 0.60$, whose characteristic depth is $\bar{z} \approx 0.16$ according to~\cite{yoon14}.

\section{Methodology} \label{method}

\subsection{The Delta-Map} \label{delta-map}

Our GNC hemispherical analysis is performed using a number counts estimator based on~\cite{alonso15}, which is defined as

\begin{equation}
\label{eq:delta_n}
\Delta_i = 2 \times \left( \frac{n_i^U - n_i^D}{n_i^U + n_i^D} \right)\;,
\end{equation}

\noindent where $n_i^j \equiv N^j_i/(4\pi f^j_{sky,i})$, $i$ denotes the hemisphere centre defined by the HEALpix pixelisation grid with $N_{side}=8$ resolution, and $j$ represents the "up" ($U$) and "down" ($D$) hemispheres according to this scheme. $N^j_i$ and $f^j_{sky,i}$ are the number of objects and the observed fraction of the sky encompassed in each of these hemispheres, respectively. We call the collection of measurements $\Delta_i$ a "delta-map''. The GNC anisotropy is calculated by extracting the dipole of the delta-map described in Eq.~(\ref{eq:delta_n}), i.e., by setting all $\{a_{\ell m}\}$ of the delta-map to zero except for those with $\ell=1$. Hence, the resulting map only contains the dipole term, whose value and position of its "hottest'' point denotes, respectively, the amplitude and direction that we will consider throughout this work\footnote{Other estimators in the literature include the quadratic $\chi^2$, similar to those adopted by~\cite{blake02}, in addition to a Shannon Entropy estimator proposed by~\cite{pandey15}. We tested the former, and found no significant discrepancy with the results provided by our delta-map method, yet much more costly in terms of computational time. The latter, however, is left for future work.}. Note, also, that there might be some non-zero couplings in the dipole term, as from the multipole, quadrupole and octopole, due to the cut-sky LSS map, but nonetheless we found it negligible in our delta-map analysis.

\subsection{Mock data} \label{mock}

The mock data used in this study are full-sky lognormal realisations of galaxy distribution created by the {\sc flask} code\footnote{\url{http://www.astro.iag.usp.br/~flask}}~\citep{xavier16}. {\sc flask} generates lognormal realizations on spherical redshift shells around the observer (i.e., tomographically) of fields described by a set of angular cross- and auto-power spectra $C_{\ell}^{ij}$ which must be given as input, where the indices $i$ and $j$ refer to the various shells. When generating galaxy distributions, the galaxies are Poisson sampled from the lognormal fields whose statistics are described by the $C_{\ell}^{ij}$s.

In this work we simulated the angular distribution of galaxies in 8 redshift shells (with top-hat profiles) equally spaced in the range $0<z<0.4$ following the WISE redshift distribution estimated by~\cite{yoon14} and the number density $C_{\ell}^{ij}$s (band-limited to $\ell_{\mathrm{max}}=512$) computed by {\sc camb sources}\footnote{\url{http://camb.info/sources}} \citep{challinor11} for galaxies with linear and constant bias $b$ under an isotropic $\Lambda$CDM model with cosmological parameters from Planck~\citep{planck15a}, and a minimal massive neutrino configuration (one massive neutrino with mass $m_\mathrm{\nu}=0.06\mathrm{eV}$). The power spectra include non-linear contributions -- modelled by {\sc halofit}~\citep{smith03, takahashi12} -- and all effects described by Eq. (30) of~\cite{challinor11} (e.g. redshift space distortions and gravitational lensing distortions of the volume elements). The boost due to our proper motion, nevertheless, has not been accounted during this procedure, yet it is incorporated afterwards as explained in more details in the next section. The galaxy distributions simulated in the shells were then projected to form a single galaxy surface density map with the same resolution as the one used for the real WISE data (given by $N_{side}=128$). The average galaxy number density in the simulations was set so as to match the observed density of $1.85 \; \times 10^{-2}\mathrm{arcmin^{-2}}$ ($\sim 14$ galaxies per pixel), and the bias $b$ was set to $1.37$ to match the variances $\sigma_{\mathrm{g}}^2$ of the galaxy counts inside the pixels ($\sigma_{\mathrm{g}}^2\simeq 45$). Finally, we applied to the simulated maps the same masking template used on the real data. An example of a simulated map is shown in the right panel of Figure~\ref{fig:wise_sim_map}. All this simulation procedure was repeated in order to generate 1000 independent mocks.      

\begin{figure*}
\includegraphics[width=6.3cm, height=8.5cm, angle=90]{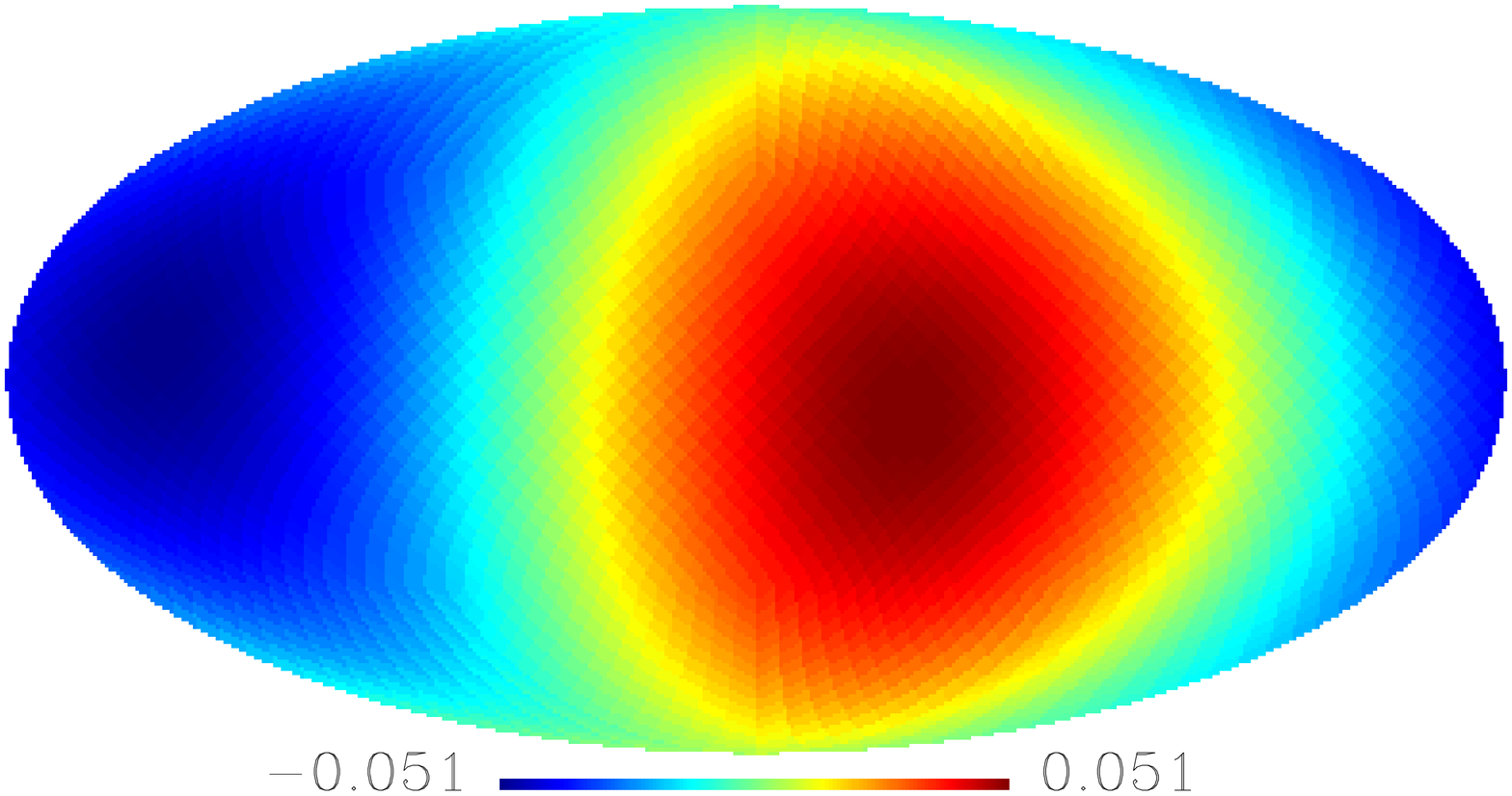}
\includegraphics[width=6.3cm, height=8.5cm, angle=90]{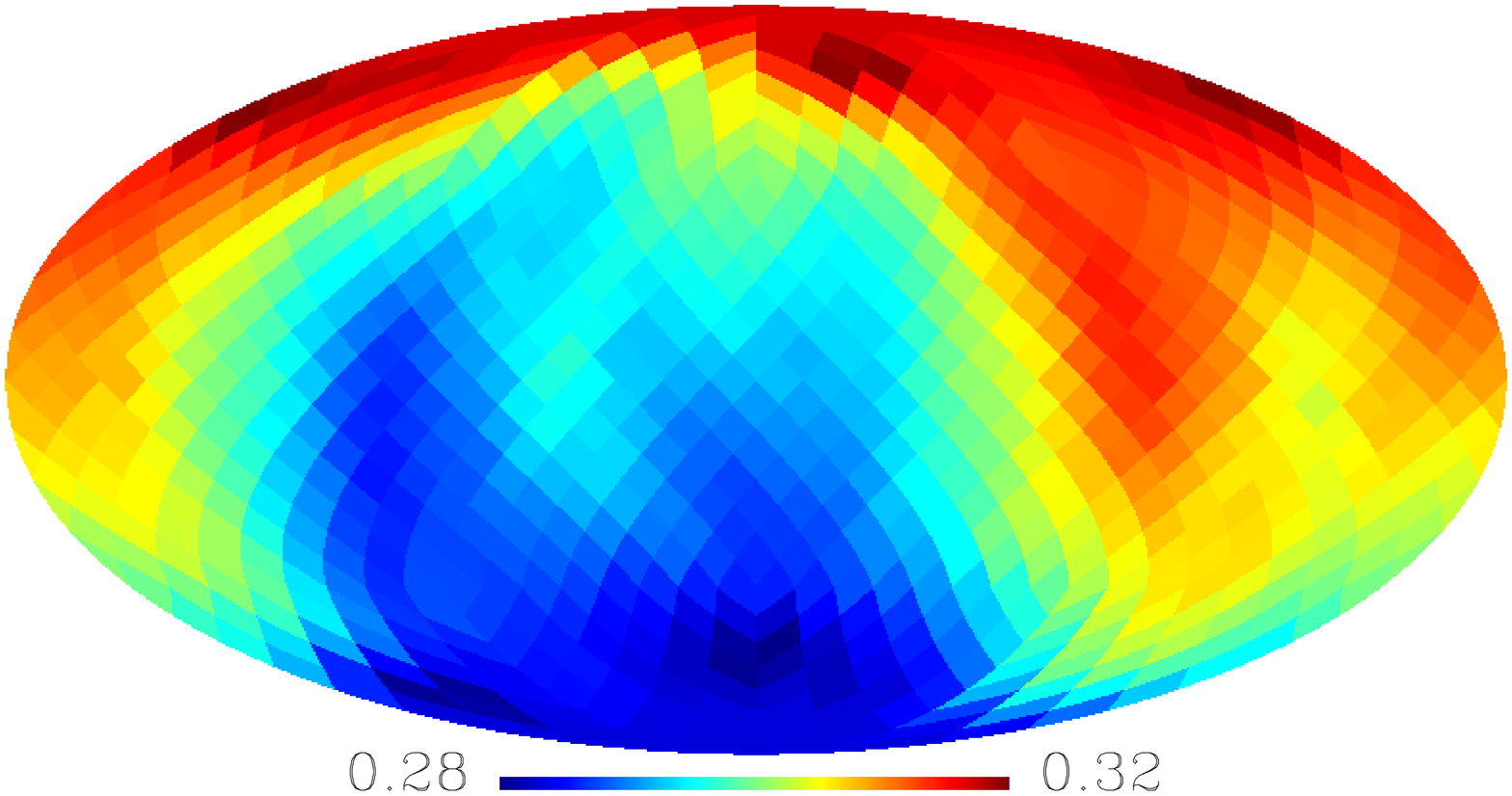}
\caption{{\it Left panel}: The dipole component obtained from the delta-map analysis of the WISE-2MASS data. We find $A = 0.0507$ for its amplitude, whose preferred direction points toward $(l,b) = (323.44^{\circ},-4.78^{\circ})$, i.e., the reddest region of the corresponding map. {\it Right panel}: The anisotropy bias introduced in the data due to the asymmetric sky coverage of the foreground mask, which has been taken into account in our analyses.}
\label{fig:wise_dipole}
\end{figure*}

\vspace{3.0cm}

\section{Results} \label{res}

\begin{table*}
\begin{tabular}{cccccc}
\hline
\hline
LSS data set & GNC dipole ($A \times 10^{-2}$) & $(l,b)$ & statistical significance & Reference\\
\hline
WISE-2MASS & $5.00$ & $(310^{\circ},-15^{\circ})$ & $0.010-0.020$ & \cite{yoon14}\\
2MPZ & - & $(315^{\circ},30^{\circ})$ & $> 0.120$  & \cite{appleby14}\\
2MPZ & $2.80$ & $(320^{\circ},6^{\circ})$ & $0.130$ & \cite{alonso15}\\
NVSS & $0.36$ & $(246^{\circ},38^{\circ})$ & $0.017$ & \cite{tiwari16}\\
WISE-2MASS & $5.07$ & $(323^{\circ},-5^{\circ})$ & $0.061$ & This work\\
\hline
\hline
SNe data set & Velocity dipole \mbox{\tt v} (km/s) & $(l,b)$ & statistical significance & Reference\\
\hline
Union2 & - & $(309^{\circ},19^{\circ})$ & $0.054$ & \cite{colin11} \\
"First Amendment" & $249$ & $(319^{\circ},7^{\circ})$ & $> 0.050$ & \cite{turnbull12} \\ 
Union2.1 & $260$ & $(295^{\circ}, 5^{\circ})$ & $\leq 0.005$ & \cite{rathaus13} \\
Union2.1 + $\mbox{SN}_{\mathrm factory}$ & $243$ & $(298^{\circ}, 15^{\circ})$ & $0.010$ & \cite{feindt13} \\
Union2.1 + 6dF + LOSS & - & $(276^{\circ}, 20^{\circ})$ & $0.290$ & \cite{appleby15} \\
\hline
\hline
SNe data set & Distance modulus dipole ($A \times 10^{-2}$) & $(l,b)$ & statistical significance & Reference\\
\hline
Union2 & - & $(309^{\circ},18^{\circ})$ & $\sim 0.333$ & \cite{antoniou10}\\
Union2 & $0.13$ & $(309^{\circ},-15^{\circ})$ & $0.048$ &\cite{mariano12}\\
Constitution & - & $(308^{\circ},-19^{\circ})$ & $> 0.005$ & \cite{kalus13}\\
Union2 & $(3.00 \pm 3.00)$ & $(306^{\circ},-13^{\circ})$ & - & \cite{cai13}\\
Union2.1 & $1.50$ & $(306^{\circ},-13^{\circ})$ & $0.076$ & \cite{bengaly15a}\\
JLA & $2.50$ & $(58^{\circ}, -60^{\circ})$ & $0.182$ & \cite{bengaly15a}\\
JLA & $< 0.20$ & $(316^{\circ},-5^{\circ})$ & - & \cite{lin16a}\\
\hline
\hline
CMB data set & Temperature dipole ($A \times 10^{-3}$) & $(l,b)$ & statistical significance & Reference\\
\hline
Planck kinematic dipole & $1.35 \pm 0.26 $ & $(264^{\circ},48^{\circ})$ & $5 \times 10^{-4}$ & \cite{planck14}\\
\hline
\hline
\end{tabular}
\caption{The amplitude, direction, and statistical significance of the dipole estimates reported in the literature from LSS and SNe with respect to this work results. The CMB kinematic dipole direction is also shown for the sake of comparison.}
\label{tab:dipoles}
\end{table*}

\subsection{WISE-2MASS data}

The result of the delta-map analysis is exhibited in the left panel of Figure~\ref{fig:wise_dipole}, where we obtained an amplitude of  $A = 0.0507$ towards the $(l,b) = (323.44^{\circ},-4.78^{\circ})$ direction. The uncertainty of this dipole amplitude can be estimated from the shot noise due to the discrete distribution of galaxies, as performed by~\cite{yoon14}. This is assessed following $\sigma_A = 1.5\, (\!\sqrt{\pi}\,)^{-1} \sqrt{\Omega/(4\pi\bar{n}})$~\citep{itoh10}, being $\Omega$ the total area spanned by the survey, and $\bar{n}$ the average number of sources {\it per} steradian. We obtained $\sigma_A = 0.0014$, which is significantly smaller than the total dipole amplitude, besides the cosmic variance, which comprises 40\% of it due to the large foreground mask adopted. This result presents good agreement with previous analyses in the literature. For instance, \cite{yoon14} obtained a maximal asymmetry with amplitude $A \simeq 0.05 \pm 0.01$ pointing roughly at the $(l,b) = (310^{\circ},-15^{\circ})$ direction using the same WISE-2MASS data, albeit a different estimator and masking procedure ($f_{sky} \simeq 0.65$), while~\cite{alonso15} analysis provided $A = 0.028$ toward $(l,b) = (320^{\circ},6^{\circ})$ with the 2MPZ data provided by a local variance estimator similar to that presented in~\cite{akrami14}. Moreover,~\cite{appleby14} found a maximal anisotropy at $(l,b) = (315^{\circ},30^{\circ})$ in the 2MPZ catalogue as well, whose estimator based on the hemispherical variance of the luminosity function instead. This may explain the larger discrepancy between their preferred direction and the aforementioned works. We remark that our results are also in relative good concordance with the anisotropy analyses carried out with SNe data, as shown in Table~\ref{tab:dipoles}, thus indicating that our directional analysis could be plausibly explained in terms of the clustering dipole arisen by the bulk flow velocity present in the low-$z$ cosmic web, whose value is commonly ascribed to our relative motion through the Universe. 

The right panel of Fig.~\ref{fig:wise_dipole}, on the other hand, features the hemispherical anisotropy of the available sky area, i.e., the $f_{sky}$ variance in antipodal portions of the celestial sphere due to the asymmetric foreground mask. The numbers presented on the colour bar correspond to the fluctuation around $0.30$, which is the average $f_{sky}$ comprised in the hemispheres\footnote{Note that $f_{sky} \simeq 0.60$ for the all sky map.}. We note that the "preferred direction" obtained in this analysis does not coincide with the GNC dipole, since we divided the total number of objects inside each of these hemispheres by this $f_{sky}$ directional variance, therefore, the incompleteness of the sky coverage does not affect the dipole estimation via delta-map in a significant manner. Moreover, we stress that this dipole signal is robust with respect to different masking procedures that presents smaller or larger $f_{sky}$, and the same happens when adopting different resolutions for the density number contrast map (as $N_{side}=64$, for instance), or different number of hemispheres (say $192$ or $3072$, corresponding to $N_{side}=4$ or $N_{side}=16$, respectively) for the delta-map analysis. When carrying out distinct magnitude cuts, the dipole amplitude slightly increases (about $10\%$ of the original $A$ value) when selecting the brightest sources (as $W_1 < 14.2$ or $W_1 < 14.5$) of the catalogue, and decreases in a similar fashion when applying an upper magnitude cut which leaves the deepest sources of the sample. This is an expected result, since the brighter objects, in general, lie closer to us, and thus are more strongly affected by the overdensities of local large-scale structure, while the opposite happens for the deeper sub-samples.  

\subsection{Statistical significance test}

\begin{figure}
\includegraphics[width=6.3cm, height=8.5cm, angle=90]{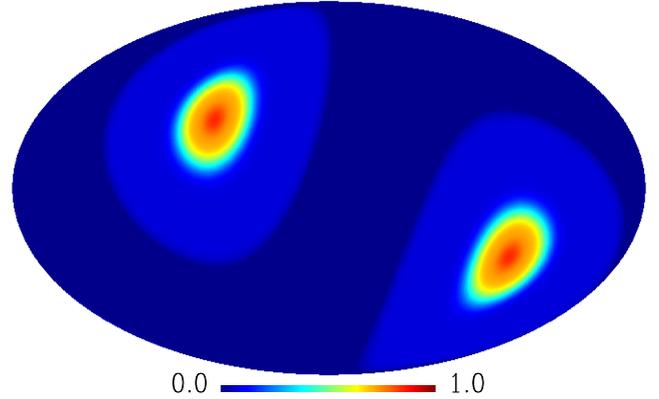}
\caption{The exposure map of the WISE satellite. The "hottest" regions of this map denotes the portions of the sky where the satellite had visited for longer periods. }
\label{fig:wise_exp}
\end{figure}

\begin{figure}
\includegraphics[height=6.3cm, width=8.5cm, angle=0]{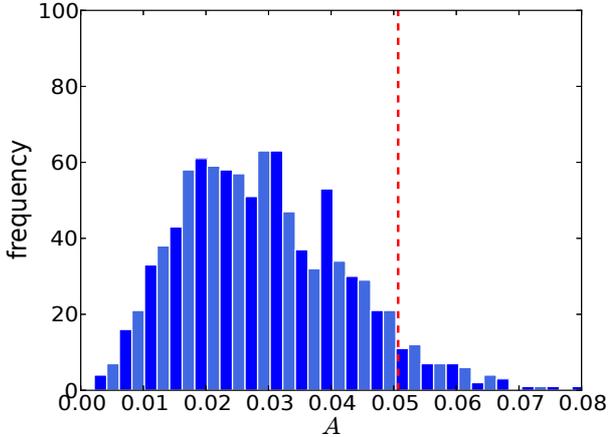}
\caption{The distribution of the 1000 log-normal realisations dipoles when the exposure map correction is taken into account. The real map dipole, $A = 0.0507$ is shown in the red vertical line with $p$-value = $0.061$.}
\label{fig:hist}
\end{figure}

We repeat the delta-map analysis of the GNC with a set of 1000 mock WISE-2MASS catalogues produced via lognormal realisations, where the underlying number density contrast field (as explained in Section~\ref{mock}) is simulated given the expected redshift and sky coverage of the sample. In addition, we develop a model for the WISE sky exposure\footnote{This sky exposure model is valid for the the main band, $W_1$, which is the most complete of all four bands available.}, as shown in Fig.~\ref{fig:wise_exp}. Since the satellite visits some patches in the sky more often than others, such as the ecliptic poles, we might expect a higher probability of objects being detected in these regions, thus leading to a potentially larger anisotropy, as found in~\cite{bengaly15b} with galaxy cluster observations. Such effect is included in the lognormal mocks by weighting the number of sources in each pixel according to this map, therefore the darkest blue regions of this map should present smaller number of objects per pixel than the hotter ones. Furthermore, the expected dipole generated by Doppler boosting and aberration of sources due to our relative motion, i.e., the actual kinematic dipole, is also incorporated in these lognormal realisations. Although the WISE-2MASS data set does not satisfy the required configuration to probe it with sufficient statistical significance~\citep{yoon15}, since this effect (of order $A \sim 10^{-3}$) is subdominant when compared to the clustering dipole in the characteristic depth of this sample ($A \sim 10^{-2}$), this signal could slightly contribute to enhance it, and hence needs to be properly accounted for. This effect is modelled as a simple dipolar modulation following $\delta N/N = (1 + A\cos{\theta})$ in the mock density contrast maps, where $\cos{\theta}$ denotes the angle between the line of sight of each pixel and the fiducial direction given according to $(l,b) = (263.99^{\circ}, 48.26^{\circ})$, i.e., the CMB dipole direction~\citep{planck14}, and the fiducial amplitude $A$ is set as $A = 0.0028$~\citep{yoon15}. 

The distribution of the dipole amplitudes $A$ obtained from these mock catalogues is shown in Fig.~\ref{fig:hist}. We find that the null hypothesis of anomalous dipolar anisotropy in the LSS is rejected, since we obtain $p$-value = $0.061$ for the actual dipole amplitude $A = 0.0507$, hence, showing no significant tension between the observed GNC dipole and the anisotropy expected from the $\Lambda$CDM matter density fluctuations. When approximating this histogram as a Gaussian distribution, we obtain a mean and standard deviation of $A = 0.0293 \pm 0.0132$, which moderately matches the expected theoretical value by~\cite{yoon14} ($A = 0.0230 \pm 0.0120$). We also tested whether the delta-map preferred direction of these realisations coincide with the real WISE-2MASS sample, finding that they are randomly oriented through the sky. This result demonstrates that no significant selection effect due to the incomplete sky coverage, or to the non-uniformities introduced by the anisotropic sky exposure, could lead to any bias in the dipole detection. Therefore, we report no statistically significance evidence for an anomalous dipole in the LSS, thus disagreeing with previous results detected in the radio sky~\citep{singal11, rubart13, tiwari14, tiwari15}, for instance, besides showing that the moderate tension between the WISE-2MASS GNC dipole and the concordance model could be reduced when the variance of the density contrast sample, as well as the satellite non-uniform sky exposure, are properly incorporated in the angular power spectrum of the matter density field.

\section{Conclusions} \label{conc}

In this work we have tested the hypothesis of cosmological isotropy in the low redshift range with the WISE-2MASS extra-galactic catalogue, which corresponds to the current largest and deepest all-sky sample in the infra-red spectrum. We have investigated whether there is agreement between the well-known dipole anisotropy in the CMB due to the imprint of our relative motion ($z \sim 1000$), with the LSS data which probes in much shallower scales ($z \sim 0.10$). We have performed color and magnitude cuts in these data in order to minimise stellar contamination and systematic biases, in addition to applying a conservative foreground mask due to dust obscurity. By means of a directional analysis based on GNC hemispherical comparison in antipodes, which is shown to be unbiased with respect to the asymmetric mask, we have obtained a dipole whose amplitude is $A = 0.0507$, pointing at the $(l,b) = (323.44^{\circ},-4.78^{\circ})$ direction, thus consistent with previous studies in the same redshift range using data of different probes and experiments. We have assessed the statistical significance of this result using WISE-2MASS mocks based on lognormal simulations, which have been produced with the {\sc FLASK} code \citep{xavier16} under the $\Lambda$CDM assumption, and we have posteriorly included observational effects such as the non-uniformities sky exposure, and a fiducial dipole modulation produced by Doppler boosting and aberration. We have found out that the GNC dipole obtained in the WISE-2MASS data is not significantly unusual with respect to these realisations ($p$-value = $0.061$). 

Thus, we conclude that there is no significant evidence for anomalous dipole anisotropy in the LSS, contrarily to suggested in previous analyses such as the moderately large GNC dipole reported in the same data set, and very large velocity dipoles detected in radio sources and galaxy clusters via kSZ as well. We note, nevertheless, that the WISE-2MASS catalogue cannot actually probe the kinematic dipole that is manifested in the CMB. As discussed in~\cite{gibelyou12} and in~\citet{yoon14}, it is expected that the clustering dipole should dominate in low redshift ranges, as in the case of the WISE-2MASS sources. It is required an observational sample comprising $N_{gal} \sim 10^7$ with $f_{sky} = 0.75$, and median redshift $\bar{z} \sim 0.70$, in order to probe such signal with $5\sigma$ confidence level, as shown in~\cite{yoon15}. Such data set with is not currently available in any redshift range or frequency observed, hence we cannot underpin that the dipole we have detected is, in fact, because of our relative motion through the Universe, the main source of anisotropy expected in the standard cosmological scenario, albeit our results is in good agreement with previous estimations of the clustering dipole (Table~\ref{tab:dipoles}). 

Finally, the amplitude of our reported dipole is also consistent with the $\Lambda$CDM GNC fluctuations obtained from its matter density power spectrum once the sample variance and WISE's celestial exposure are properly taken into account. Nevertheless, a mild discrepancy between this dipole direction between the local probes and the CMB kinematic dipole still persists. This could be ascribed to the fact that the WISE-2MASS sample is not deep enough to obtain the expected convergence between them, as discussed in~\cite{gibelyou12}, or perhaps there are some unaccounted anisotropy introduced by the presence of very large structures, such as the Sloan Great Wall~\citep{gott05} or the Eridanus Supervoid~\citep{nadathur14, szapudi15, kovacs15b, finelli16}. A more detailed assessment of such structures in the GNC anisotropy study has yet to be performed, although~\cite{rubart14, bolejko16} showed that large underdensities are unable to induce such large LSS dipoles (albeit~\cite{kraljic16} disputed this result). Our final conclusion is that one of the foundations of the concordance cosmology, i.e., the cosmic isotropy, is an assumption that is actually consistent with modern astrophysical data, given their current limitations. However, the prospect of probing the large-scale isotropy should be tremendously improved with the advent of the next-generation LSS surveys such as Large Synoptic Sky Survey (LSST)~\citep{lsst}, which has been shown to be capable of probing the cosmic dipole with much better precision than current surveys~\citep{itoh10}, and especially with Square Kilometer Array (SKA)~\citep{ska1, ska2, ska3}, since this experiment should provide even larger data sets covering a wide area of the sky ($f_{sky} \simeq 0.75$). All these efforts will enable to put under scrutiny the crucial assumptions of the standard cosmological scenario with unprecedented precision.

\section*{acknowledgments}

The authors thank CNPq, CAPES, INEspa\c{c}o,  FAPERJ and FAPESP for the grants under which this work was carried out. CPN is also supported by the DTI-PCI  program of the Brazilian Ministry of Science, Technology and Innovation (MCTI). AB  acknowledges the {\em Science without Borders Program} of CAPES for a PVE project (88881.064966/2014-01). The authors also acknowledge the HEALpix package for the derivation of the analyses performed in this work. In addition, this research has made use of the NASA/ IPAC Infrared Science Archive, which is operated by the Jet Propulsion Laboratory, California Institute of Technology, under contract with the National Aeronautics and Space Administration, and we acknowledge the use of the Legacy Archive for Microwave Background Data Analysis (LAMBDA), part of the High Energy Astrophysics Science Archive Center (HEASARC). HEASARC/LAMBDA is a service of the Astrophysics Science Division at the NASA Goddard Space Flight Center. JSA thanks IIP-UFRN for the hospitality where this work was completed.

\bsp	
\label{lastpage}


\begin{thebibliography}{99}

\bibitem[\protect\citeauthoryear{Abell et al.}{2009}]{lsst}
Abell, P. A. {\it et al.} [LSST Red Book 2.0], 2009, arXiv:0912.0201
%
\bibitem[\protect\citeauthoryear{Ade et al.}{2015a}]{planck15a} 
Ade, P. A. R. {\it et al.} [Planck collaboration], 2015, arXiv:1502.01589
%
\bibitem[\protect\citeauthoryear{Ade et al.}{2015b}]{planck15b} 
Ade, P. A. R. {\it et al.} [Planck collaboration], 2015, arXiv:1506.07135
%
\bibitem[\protect\citeauthoryear{Aghanim et al.}{2014}]{planck14} 
Aghanim, N. {\it et al.}, 2014, Astron. Astrophys. 571, A27
%
\bibitem[\protect\citeauthoryear{Akrami et al.}{2014}]{akrami14}
Akrami, Y. {\it et al.}, 2014, Astrophys. J., 784, L42
%
\bibitem[\protect\citeauthoryear{Alcaniz \& Lima}{1999}]{alcaniz99}
Alcaniz, J. S. \& Lima, J. A. S., 1999, Astrophys.\ J.\  { 521},  L87
%
\bibitem[\protect\citeauthoryear{Alcaniz et al.}{2003}]{alcaniz03}
Alcaniz, J. S. {\it et al.}, 2003, Mon.\ Not.\ Roy.\ Astron.\ Soc.\  {340}, L39
%
\bibitem[\protect\citeauthoryear{Alonso et al.}{2015}]{alonso15} 
Alonso, D. {\it et al.}, 2015, Mon. Not. Royal Ast. Soc., 449, 670
%
\bibitem[\protect\citeauthoryear{Antoniou \& Perivolaropoulos}{2010}]{antoniou10}
Antoniou, I. \& Perivolaropoulos, L., 2010, JCAP, 12, 012
%
\bibitem[\protect\citeauthoryear{Appleby \& Shafieloo}{2014}]{appleby14} 
Appleby. S. \& Shafieloo, A., 2014, JCAP, 10, 070
%
\bibitem[\protect\citeauthoryear{Appleby et al.}{2015}]{appleby15}
Appleby, S. {\it et al.} 2015, Astrophys. J., 801, 2, 76 
%
\bibitem[\protect\citeauthoryear{Atrio-Barandela et al.}{2015}]{atrio15}
Atrio-Barandela, F. {\it et al.}, 2015, Astrophys. J., 810, 2
%
\bibitem[\protect\citeauthoryear{Aubourg et al.}{2015}]{boss15} 
Aubourg, E. {\it et al.} [BOSS colaboration], 2015, Phys. Rev. D, 92, 123516
%
\bibitem[\protect\citeauthoryear{Baleisis et al.}{1998}]{baleisis98} 
Baleisis, A. {\it et al.}, 1998, Mon. Not. Royal Ast. Soc., 297, 545
%
\bibitem[\protect\citeauthoryear{Basilakos \& Plionis}{1998}]{basilakos98} 
Basilakos, S. \& Plionis, M., 1998, Mon. Not. Royal Ast. Soc., 299, 637
%
\bibitem[\protect\citeauthoryear{Basilakos \& Plionis}{2006}]{basilakos06} 
Basilakos, S. \& Plionis, M., 2006, Mon. Not. Royal Ast. Soc., 373, 1112
%
\bibitem[\protect\citeauthoryear{Bengaly et al.}{2015a}]{bengaly15a}
Bengaly, C. A. P. {\it et al.}, 2015a, Astrophys. J., 808, 39
%
\bibitem[\protect\citeauthoryear{Bengaly et al.}{2015b}]{bengaly15b}
Bengaly, C. A. P. {\it et al.}, 2015b, arXiv:1511.09414
%
\bibitem[\protect\citeauthoryear{Bernui}{2008}]{b08} 
Bernui, A., 2008, Phys. Rev. D, 78, 063531 
%
\bibitem[\protect\citeauthoryear{Bernui et al.}{2014}]{bop14}
Bernui, A. {\it et al.}, 2014, JCAP, 10, 041 
%
\bibitem[\protect\citeauthoryear{Betoule et al.}{2014}]{betoule14} 
Betoule, M. {\it et al.}, 2014, Astron. Astrophys., 568, A22
%
\bibitem[\protect\citeauthoryear{Beutler et al.}{2011}]{beutler11} 
Betoule, M. {\it et al.}, 2011, Mon. Not. Roy. Astron. Soc., 416, 3017
%
\bibitem[\protect\citeauthoryear{Bilicki et al.}{2011}]{bilicki11}
Bilicki, M. {\it et al.}, 2011, Astrophys. J., 741, 31
%
\bibitem[\protect\citeauthoryear{Bilicki et al.}{2014}]{bilicki14}
Bilicki, M. {\it et al.}, 2014, Astrophys. J. Suppl. Ser., 2010, 9
%
\bibitem[\protect\citeauthoryear{Blake \& Wall}{2002}]{blake02} 
Blake, C. \& Wall, J., 2002, Nature, 416, 150
%
\bibitem[\protect\citeauthoryear{Blake et al.}{2011a}]{blake11a} 
Blake, C. {\it et al.}, 2011, Mon. Not. Roy. Astron. Soc., 415, 2876
%
\bibitem[\protect\citeauthoryear{Blake et al.}{2011b}]{blake11b} 
Blake, C. {\it et al.}, 2011, Mon. Not. Roy. Astron. Soc., 418, 1707
%
\bibitem[\protect\citeauthoryear{Bolejko et al.}{2016}]{bolejko16}
Bolejko, K. {\it et al.}, 2016, JCAP, 06, 035 
%
\bibitem[\protect\citeauthoryear{Boughn et al.}{2002}]{boughn02}
Boughn, S. P. {\it et al.}, 2002 Astrophys. J., 580, 672
%
\bibitem[\protect\citeauthoryear{Cai et al.}{2013}]{cai13} 
Cai, R. G. {\it et al.}, 2013, Phys. Rev. D 87, 123522
%
\bibitem[\protect\citeauthoryear{Challinor \& Lewis}{Challinor \& Lewis}{2011}]{challinor11}
Challinor A.,  Lewis A.,  2011, Phys. Rev. D, 84, 043516
%
\bibitem[\protect\citeauthoryear{Cheng et al.}{2016}]{cheng16}
Cheng, C., Zhao, W., Huang, Q.-G., 2016, Phys. Lett. B, 757, 445 
%
\bibitem[\protect\citeauthoryear{Carvalho \& Marques}{2015}]{carvalho15}
Carvalho, C. S. \& Marques, M., 2015, arXiv:1512.07869
%
\bibitem[\protect\citeauthoryear{Carvalho \& Basilakos}{2016}]{carvalho16}
Carvalho, C. S. \& Basilakos, S., 2016, arXiv:1603.07519
%
\bibitem[\protect\citeauthoryear{Carvalho et al.}{2016}]{gabriela16}
Carvalho, G. C. {\it et al.}, 2016, Phys. Rev. D, 93 no.2, 023530
%
\bibitem[\protect\citeauthoryear{Clarkson}{2012}]{clarkson12}
Clarkson, C., 2012, Comp. Rend. de l'Acad des Sci., 13, 682
%
\bibitem[\protect\citeauthoryear{Colin et al.}{2011}]{colin11}
Colin, J., Mohayee, R., Sarkar, S., Shafieloo, A., 2011, Mon. Not. Royal Ast. Soc., 414, 264
%
\bibitem[\protect\citeauthoryear{Conklin}{1969}]{conklin69}
Conklin, E. K. 1969, Nature, 222, 971
%
\bibitem[\protect\citeauthoryear{Dai et al.}{2011}]{dai11}
Dai, D.-C. {\it et al.}, 2011, JCAP, 04 015
%
\bibitem[\protect\citeauthoryear{De Vaucouleurs \& Peters}{1968}]{devaucouleurs68}
De Vaucouleurs, G. \& Peters, W. L., 1968, Nature, 220, 868
%
\bibitem[\protect\citeauthoryear{Eisenstein et al.}{2005}]{eisenstein05}
Eisenstein, D. J. {\it et al.} [SDSS Collaboration], 2005, Astrophys. J., 633, 560
%
\bibitem[\protect\citeauthoryear{Erdo\v{g}du et al.}{2006}]{erdogdu06}
Erdo\v{g}du, P. {\it et al.}, 2006, Mon. Not. Royal Ast. Soc., 368, 1515
%
\bibitem[\protect\citeauthoryear{Eriksen et al.}{2004}]{eriksen04}
Eriksen, H. K. {\it et al.}, 2004, Astrophys. J., 609, 1198
%
\bibitem[\protect\citeauthoryear{Feindt et al.}{2013}]{feindt13}
Feindt, U. {\it et al.}, 2013, Astron. Astrophys. 560, A90
%
\bibitem[\protect\citeauthoryear{Fern\'andez-Cobos et al.}{2014}]{cobos14}
Fern\'andez-Cobos, R. {\it et al.}, 2014, Mon. Not. Roy. Astron. Soc. 441, 2392
%
\bibitem[\protect\citeauthoryear{Finelli et al.}{2016}]{finelli16}
Finelli, F. {\it et al.}, 2016, Mon. Not. Roy. Astron. Soc. 455, 1246
%
\bibitem[\protect\citeauthoryear{Goodman}{1995}]{goodman95}
Goodman, J., 1995, Phys. Rev. D, 52, 1821
%
\bibitem[\protect\citeauthoryear{G\'orski et al.}{2005}]{gorski05}
G\'orski, K. M. {\it et al.}, 2005, Astrophys. J., 622, 759
%
\bibitem[\protect\citeauthoryear{Gott III et al.}{2005}]{gott05}
Gott III, J. R. {\it et al.}, 2005, Astrophys. J. 624 (2005) 463
%
\bibitem[\protect\citeauthoryear{Ghosh et al.}{2016}]{ghosh16}
Ghosh, S. {\it et al.}, 2016, JCAP, 01, 046
%
\bibitem[\protect\citeauthoryear{Gibelyou \& Huterer}{2012}]{gibelyou12}
Gibelyou, C. \& Huterer, D., 2012, Mon. Not. Royal Ast. Soc., 427, 1994 
%
\bibitem[\protect\citeauthoryear{Hansen et al.}{2004}]{hansen04}
Hansen, F. K. {\it et al.}, 2004, Mon. Not. Royal Ast. Soc., 354, 641
%
\bibitem[\protect\citeauthoryear{Hazra \& Shafieloo}{2015}]{hazra15}
Hazra, D. J. \& Shafieloo, A., 2015, JCAP, 11, 012
%
\bibitem[\protect\citeauthoryear{Hinshaw et al.}{2009}]{hinshaw09}
Hinshaw, G., 2009, Astrophys. J. Suppl. 180, 225
%
\bibitem[\protect\citeauthoryear{Hinshaw et al.}{2013}]{hinshaw13}
Hinshaw, G. {\it et al.} [WMAP Collaboration], 2013, Astrophys.\ J.\ Suppl.\  {208}, 19
%
\bibitem[\protect\citeauthoryear{Hoftuft et al.}{2009}]{hoftuft09}
Hoftuft, J. {\it et al.}, 2009, Astrophys. J., 699, 985
%
\bibitem[\protect\citeauthoryear{Hong et al.}{2014}]{hong14}
Hong, T. {\it et al.}, 2014, Mon. Not. Royal Ast. Soc., 445, 402
%
\bibitem[\protect\citeauthoryear{Itoh et al.}{2010}]{itoh10}
Itoh, Y., Yahata, K., Takada, M., 2010, Phys. Rev. D., 82, 043530
%
\bibitem[\protect\citeauthoryear{Jarvis et al.}{2015}]{ska1}
Jarvis, M. {\it et al.}, 2015, arXiv:1501.03825
%
\bibitem[\protect\citeauthoryear{Javanmardi et al.}{2015}]{javanmardi15}
Javanmardi, B. {\it et al.}, 2015, Astrophys. J., 810, 47
%
\bibitem[\protect\citeauthoryear{Jim\'enez et al.}{2015}]{jimenez15}
Jim\'enez, J. N. {\it et al.}, 2015, Phys. Lett. B, 741, 168
%
\bibitem[\protect\citeauthoryear{Kalus et al.}{2013}]{kalus13}
Kalus, N. {\it et al.}, 2013, Astron. Astrophys., 513, A56
%
\bibitem[\protect\citeauthoryear{Kashlinsky et al.}{2009}]{kashlinsky09}
Kashlinsky, A. {\it et al.}, 2009, Astrophys. J., 686, L49
%
\bibitem[\protect\citeauthoryear{Kashlinsky et al.}{2010}]{kashlinsky10}
Kashlinsky, A. {\it et al.}, 2010, Astrophys. J., 712, L81
%
\bibitem[\protect\citeauthoryear{Kashlinsky et al.}{2011}]{kashlinsky11}
Kashlinsky, A. {\it et al.}, 2011, Astrophys. J., 732, 1
%
\bibitem[\protect\citeauthoryear{Kocevski et al.}{2004}]{kocevski04}
Kocevski, D. D. {\it et al.}, 2004, Astrophys. J., 608, 721
%
\bibitem[\protect\citeauthoryear{Kogut et al.}{2013}]{kogut93}
Kogut, A. {\it et al.}, 1993, Astrophys. J., 419, 1
%
\bibitem[\protect\citeauthoryear{Kov\'acs \& Szapudi}{2015}]{kovacs15a}
Kov\'acs, A. \& Szapudi, I., 2015, Mon. Not. Roy. Astron. Soc. 448, 2, 1305
%
\bibitem[\protect\citeauthoryear{Kov\'acs \& Garc\'ia-Bellido}{2015}]{kovacs15b}
Kov\'acs, A. \& Garc\'ia-Bellido, J., 2015, arXiv:1511.09008 
%
\bibitem[\protect\citeauthoryear{Kraljic \& Sarkar}{2016}]{kraljic16}
Kraljic, D. \& Sarkar, S., 2016, arXiv:1607.07377 
 %
\bibitem[\protect\citeauthoryear{Lahav et al.}{1988}]{lahav88} 
Lahav, O. {\it et al.}, 1988, Mon. Not. Royal Ast. Soc., 234, 677
%
\bibitem[\protect\citeauthoryear{Lin et al.}{2016a}]{lin16a} 
Lin, H. N. {\it et al.}, 2016, Mon. Not. Royal Ast. Soc., 456, 1881
%
\bibitem[\protect\citeauthoryear{Lin et al.}{2016b}]{lin16b}
Lin, H. N. {\it et al.}, 2016, arXiv:1604.07505  
%
\bibitem[\protect\citeauthoryear{Lynden-Bell et al.}{1989}]{lynden-bell89} 
Lynden-Bell, D., Lahav, O., Burstein, D. 1989, Mon. Not. Royal Ast. Soc., 241, 325
%
\bibitem[\protect\citeauthoryear{Maartens}{2011}]{maartens11} 
Maartens, R., 2011, Phil. Trans. R. Soc. A, 369, 5115 
%
\bibitem[\protect\citeauthoryear{Maartens et al.}{2015}]{ska2}
Maartens, R. {\it et al.}, 2015, arXiv:1501.04076
%
\bibitem[\protect\citeauthoryear{Ma \& Pan}{2013}]{ma13}
Ma, Y.-Z. \& Pan, J., 2013, Mon. Not. Royal Ast. Soc., 437,1996
%
\bibitem[\protect\citeauthoryear{Maller et al.}{2003}]{maller03}
Maller, A. H. {\it et al.}, 2003, Astrophys. J., 598, L1
%
\bibitem[\protect\citeauthoryear{Mariano \& Perivolaropoulos}{2012}]{mariano12}
Mariano, A. \& Perivolaropoulos, L., 2012, Phys. Rev. D, 86, 083517
%
\bibitem[\protect\citeauthoryear{McKay \& Wiltshire}{2016}]{mckay16}
McKay, J. H. \& Wiltshire, D. L., 2016, Mon. Not. Roy. Astron. Soc. 457, 3285
%
\bibitem[\protect\citeauthoryear{Migkas \& Plionis}{2016}]{migkas16}
Migkas, T. \& Plionis, M., 2016, Rev. Mex. Astron. Astrofis., 52, 133
%
\bibitem[\protect\citeauthoryear{Nadathur et al.}{2014}]{nadathur14}
Nadathur, S. {\it et al.}, 2014, Phys. Rev. D, 90, 103510
%
\bibitem[\protect\citeauthoryear{Pandey}{2015}]{pandey15}
Pandey, B. 2015, arXiv:1512.03562
%
\bibitem[\protect\citeauthoryear{Peebles}{1980}]{peebles80}
Peebles, P. J. E., Large-scale Structure of the Universe, 1980, Princeton University Press
%
\bibitem[\protect\citeauthoryear{Peacock et al.}{2001}]{peacock01}
Peacock, J. A. {\it et al.} [2dFGRS experiment], 2001, Nature, 410,169
%
\bibitem[\protect\citeauthoryear{Perlmutter et al.}{1999}]{p99}
Perlmutter, S. {\it et al.} [Supernova Cosmology Project Collaboration], 1999,  Astrophys.\ J.\  {\bf 517}, 565
%
\bibitem[\protect\citeauthoryear{Plionis \& Kolokotronis}{1998}]{plionis98}
Plionis, M. \& Kolokotronis, V., 1998, Astrophys. J., 500, 1
%
\bibitem[\protect\citeauthoryear{Plionis \& Georgantopoulos}{1999}]{plionis99}
Plionis, M. \& Georgantopoulos, I., 1999, Mon. Not. Royal Ast. Soc., 306, 112 
%
\bibitem[\protect\citeauthoryear{Polastri et al.}{2015}]{polastri15}
Polastri, L., Gruppuso, A., Natoli, P., 2015, JCAP, 04, 018
%
\bibitem[\protect\citeauthoryear{Pullen \& Hirata}{2010}]{pullen10}
Pullen, A. R. \& Hirata, C. M., 2010, JCAP, 05, 027
%
\bibitem[\protect\citeauthoryear{Rathaus et al.}{2013}]{rathaus13} 
Rathaus, B., Kovetz, E. D., Itzhaki, N., 2013, Mon. Not. Royal Ast. Soc., 431, 3678
%
\bibitem[\protect\citeauthoryear{Riess et al.}{1998}]{riess98}
Riess, A. G. {\it et al.} [Supernova Search Team Collaboration], 1998, Astron.\ J.\  {\bf 116}, 1009
%
\bibitem[\protect\citeauthoryear{Reid et al.}{2012}]{reid12}
Reid, B. A. [BOSS collaboration], 2012, Mon. Not. Roy. Astron. Soc., 426, 2719
%
\bibitem[\protect\citeauthoryear{Rowan-Robinson et al.}{2000}]{rowan-robinson00} 
Rowan-Robinson, M. {\it et al.}, 2000, Mon. Not. Royal Ast. Soc., 314, 375
%
\bibitem[\protect\citeauthoryear{Rubart \& Schwarz}{2013}]{rubart13}
Rubart, M. \& Schwarz, D. J., 2013, Astron. Astrophys., 555, A117
%
\bibitem[\protect\citeauthoryear{Rubart et al.}{2014}]{rubart14}
Rubart, M., Bacon, D. \& Schwarz, D. J., 2014, Astron. Astrophys., 565, A111
%
\bibitem[\protect\citeauthoryear{Scharf et al.}{2000}]{scharf00}
Scharf, C. A. {\it et al.}, 2000, Astrophys. J., 544, 49
%
\bibitem[\protect\citeauthoryear{Schlegel et al.}{1998}]{schlegel98}
Schlegel, D. J., Finkbeiner, D. P., Davis, M., 1998, Astrophys. J., 500, 525
%
\bibitem[\protect\citeauthoryear{Schwarz et al.}{2015a}]{ska3}
Schwarz, D. J. {\it et al.}, 2015a, arXiv:1501.03820 
%
\bibitem[\protect\citeauthoryear{Schwarz et al.}{2015b}]{schwarz15}
Schwarz, D. J. {\it et al.}, 2015b, arXiv:1510.07929
%
\bibitem[\protect\citeauthoryear{Singal}{2011}]{singal11}
Singal, A., 2011, Astrophys. J., 722, L23
%
\bibitem[\protect\citeauthoryear{Smith et al.}{2003}]{smith03}
Smith R. E. {\it et al.}, 2003, Mon. Not. Roy. Astron. Soc., 341, 1311
%
\bibitem[\protect\citeauthoryear{Springbob}{2016}]{springbob16}
Springbob, C. M. {\it et al.}, 2016, Mon. Not. Roy. Astron. Soc., 456, 1886
%
\bibitem[\protect\citeauthoryear{Stewart \& Sciana}{1967}]{stewart67}
Stewart, J. M. \& Sciama, D. W., 1967, Nature, 216, 748
%
\bibitem[\protect\citeauthoryear{Strauss et al.}{1992}]{strauss92}
Strauss, M. A. {\it et al.}, 1992, Astrophys. J., 397, 395
%
\bibitem[\protect\citeauthoryear{Suzuki et al.}{2012}]{suzuki12}
Suzuki, N. {\it et al.}, 2012, Astroph. J., 746, 85
%
\bibitem[\protect\citeauthoryear{Szapudi et al.}{2015}]{szapudi15}
Szapudi, I. {\it et al.}, 2015, Mon. Not. Roy. Astron. Soc., 450, 288
%
\bibitem[\protect\citeauthoryear{Takahashi et al.}{2012}]{takahashi12}
Takahashi R. {\it et al.}, 2012, Astrophys. J., 761, 152
%
\bibitem[\protect\citeauthoryear{Tiwari et al.}{2014}]{tiwari14}
Tiwari, P. {\it et al.}, 2014, Astropart. Phys., 61, 1
%
\bibitem[\protect\citeauthoryear{Tiwari \& Jain}{2015}]{tiwari15}
Tiwari, P. \& Jain, P., 2015, Mon. Not. Roy. Astron. Soc., 447, 2658
%
\bibitem[\protect\citeauthoryear{Tiwari \& Nusser}{2016}]{tiwari16}
Tiwari, P. \& Nusser, A., 2016, JCAP, 16, 03 062 
%
\bibitem[\protect\citeauthoryear{Turnbull et al.}{2012}]{turnbull12}
Turnbull, S. J. {\it et al.}, 2012, Mon. Not. Roy. Astron. Soc. 420, 447
%
\bibitem[\protect\citeauthoryear{Watkins et al.}{2009}]{watkins09}
Watkins, R., {\it et al.}, 2009, Mon. Not. Roy. Astron. Soc., 392, 743
%
\bibitem[\protect\citeauthoryear{Wiltshire et al.}{2013}]{wiltshire13}
Wiltshire, D. L. {\it et al.}, 2013, Phys. Rev. D., 88, 083529
%
\bibitem[\protect\citeauthoryear{Wright et al.}{2010}]{wright10}
Wright, E. {\it et al.}, 2010, The Astron. J., 140, 1868
%
\bibitem[\protect\citeauthoryear{Wu et al.}{1999}]{wu99}
Wu, K. K. S. {\it et al.}, 1999, Nature, 397, 225
%
\bibitem[\protect\citeauthoryear{Xavier et al.}{2016}]{xavier16}
Xavier, H. S. {\it et al.}, 2016, Mon. Not. Roy. Astron. Soc., 459, 3693
%
\bibitem[\protect\citeauthoryear{Yahil et al.}{1986}]{yahil86}
Yahil, A., Walker, X., Rowan-Robinson, M. 1986, Astrophys. J., 301, L1
%
\bibitem[\protect\citeauthoryear{Yoon et al.}{2014}]{yoon14}
Yoon, M. {\it et al.}, 2014, Mon. Not. Royal Ast. Soc. 445, L60
%
\bibitem[\protect\citeauthoryear{Yoon \& Huterer}{2015}]{yoon15}
Yoon, M. \& Huterer, D., 2015, Astrophys. J. 813, L18 
%
\bibitem[\protect\citeauthoryear{Zhao \& Santos}{2015}]{zhao15}
Zhao, W. \& Santos, L., 2015, The Universe, 3, 9

\end{thebibliography}
\end{document}